\documentclass[sigconf]{acmart}

\usepackage{subcaption}
\usepackage{graphicx}

\AtBeginDocument{%
  }

\setcopyright{none}
\copyrightyear{2025}
\acmYear{2025}
\acmConference[XAIxArts 2025]{Explainable AI for the Arts Workshop 2025}{June 23, 2025}{Online}

\begin{document}

\title[Audiences Seek Artistic Value in Generative AI's Contributions]{What's Behind the Magic? Audiences Seek Artistic Value in Generative AI's Contributions to a Live Dance Performance}

\author{Jacqueline Elise Bruen}
\affiliation{%
  \institution{Virginia Tech}
  \city{Blacksburg}
  \state{Virginia}
  \country{USA}}
\email{jacquelineb@vt.edu}

\author{Myounghoon Jeon}
\affiliation{%
  \institution{Virginia Tech}
  \city{Blacksburg}
  \state{Virginia}
  \country{USA}}
\email{myounghoonjeon@vt.edu}

\renewcommand{\shortauthors}{Bruen and Jeon}

\begin{abstract}
 With the development of generative artificial intelligence (GenAI) tools to create art, stakeholders cannot come to an agreement on the value of these works. In this study we uncovered the mixed opinions surrounding art made by AI.

We developed two versions of a dance performance augmented by technology either with or without GenAI. For each version we informed audiences of the performance's development either before or after a survey on their perceptions of the performance. There were thirty-nine participants (13 males, 26 female) divided between the four performances. Results demonstrated that individuals were more inclined to attribute artistic merit to works made by GenAI when they were unaware of its use. We present this case study as a call to address the importance of utilizing the social context and the users' interpretations of GenAI in shaping a technical explanation, leading to a greater discussion that can bridge gaps in understanding.

\end{abstract}


\maketitle






\section{Introduction}
While the use of computers to help create art has been accepted by the general public, GenAI has brought this into question; GenAI can produce high-quality artistic media \cite{epstein2023art, feuerriegel2024generative, ibm}. While research has been done on how GenAI might impact society \cite{spennemann2024generative, karpouzis2024plato, fui2023generative}, and individuals have written commentary on how it can either be a boon or bane to the art community \cite{henriksen2024generative, jones2024visual, doi:10.1126/science.adh4451, AIartand, ifitwasnt}, little research has been done to uncover the biases against GenAI art and how these biases shape viewers' reception of explanations of how GenAI art is made. The present study seeks to understand audience perception when we (RQ1) withhold information on how a technologically augmented artwork was made, and (RQ2) alter the type of technology used (GenAI versus traditional digital tools).

We developed two versions of a technologically augmented dance performance; one made using AI software and one without. We developed a Likert scale survey for audience members which included questions on the creative value of the performances' technology. To uncover potential biases, we withheld information on how the performances were made to half of our audience members until after the survey. Hence, in this 2x2 between-subjects design we held four different performances (AI/Tell Before, AI/Tell After, Non-AI/Tell Before, Non-AI/Tell After). We analyzed the results of the Likert scale \cite{likert1932technique} surveys using Mann-Whitney tests \cite{mann1947test, wilcoxon1992individual} and found trends between different pairs of performance responses outlined in Table \ref{table:stats}. 

The present work has both practical and theoretical implications. From a practical perspective, artists can use our findings to inform how they make and present their work. Understanding how subjective works are presented to and received by audiences can inform fields that apply art and technology, such as advertising. From a theoretical perspective, our work investigates  people's value judgments of art which incorporates different types of technology. These judgments can shape how we approach explaining AI, perhaps by emphasizing elucidations on specific areas of concern for those with biases. Our investigation highlights how an art viewer's beliefs can guide the XAI community towards explaining GenAI in an arts context.

\section{Methods}

\subsection{Participants}
For the two "Non-AI" performances, we used technology to alter the visuals and sound. Participants heard about technology either before the performance or after the performance and survey. Eight participants attended the Non-AI/Tell Before performance, (3 male, 5 female) with an average age of 23.75 (SD = 4.80). Ten participants attended the Non-AI/Tell After performance, (4 male, 6 female) with an average age of 25.30 (SD = 4.99).

In the "AI" performances, we incorporated AI technology. Participants were told about the technology in a similar fashion to the Non-AI performances. Twelve participants attended the AI/Tell Before performance, (4 male, 8 female) with an average age of 29 (SD = 7.11). Nine participants attended the AI/Tell After performance, (2 male, 7 female) with an average age of 23.89 (SD = 3.89).

\subsection{Materials}
We developed two versions of a performance with a professional dancer along with a survey. Figures 1 and 2 in Appendix \ref{ase:appendixphotos} showcase the system architectures for each performance, and Figures 3 and 4 in the same appendix depict our development process. We used the Vernier Go Direct Respiration Belt and the Qualisys Track Manager (QTM) to obtain live data of the dancers' breathing and location during the performances. The Qualisys infrared cameras detected the reflective markers we positioned on the dancers' ankles. Live data were used to change the visuals on a large, mounted television screen behind the dancer and create a sonification for scene 2. 

The dancer's choreography and the performance's structure were kept the same in the two versions. The performances differed in the use of technology. In the non-AI performances, the creative technological decisions were made by the technologist. In the AI performances, some of the creative portions of the technologist's work were made or supported by AI. The two versions of the 12 minute performance included three, four minute scenes. In scene 1, petals drifted through the sky with the dancer's respiratory rate. The scene was silent. Figure 1a in Appendix \ref{ase:appendixphotos} depicts the AI imagery. In scene 2, waves moved with the dancer's location. Sonification of the dancer's position was played. Figure 1b in Appendix \ref{ase:appendixphotos} depicts the AI imagery. In scene 3, stars brightened and multiplied with the dancer's breathing patterns. Figure 1c in Appendix \ref{ase:appendixphotos} depicts the AI imagery. Music was chosen by the dancer.

In the Non-AI performance version, decisions on how to map data to visuals and sound were made by the technologist. The technologist sourced imagery from human-produced visual works, which were then made to dynamically change with the live data by layering multiple images and altering them conditionally according to data mappings designed by the technologist. 

The data mapping for the petals in scene 1 increased the distance the petals moved from their last position and brightened the sky in the background when the dancer's breathing rate increased. The mapping for the waves in scene 2 moved the waves across three axes in accordance with the dancer's position in the room. The mapping for the stars in scene 3 increased the brightness of the stars as the dancer's breathing rate increased. Figure 6 in Appendix \ref{ase:appendixphotos} shows the dancer during the performance.

The data mapping for the sonification was made by segmenting the stage into five equal sections along each of the three axes (x, y, and z). Each axis was mapped to a different musical scale. At any location where the dancer was, the audience would hear three notes (one note from each of the three axis scales) played in unison.

In the AI performance version, the technologist delegated the imagery and data mapping designs to AI (ChatGPT) \cite{openai2024chatgpt}, and in scene 2, three neural network models were built using three dance subroutines. Models were built using \citeauthor{fiebrink2010wekinator}'s Wekinator \cite{fiebrink2010wekinator}. Each neural network had four connected inputs, one hidden layer, and four nodes per hidden layer. The four inputs to train each model were the x and y coordinate sums for each of the four reflective markers attached to the dancer's ankles (two markers per ankle). The neural networks provided probabilities of each of these dance subroutines currently happening during the live performance. When asking the GenAI for data-mapping ideas, it produced many options. The technologist consulted the dancer to come to a decision. Our prompts to ChatGPT and the responses we used are in the Appendix \ref{ase:prompts}. 

The survey (in the Appendix \ref{ase:audiencesurvey}) collected audiences' feedback on the performances. The survey consisted of 29 five point Likert scale questions for the "Tell Before" performances. For the two "Tell After" performances, participants could also add any data-mapping connections they may have noticed. The survey took 10 minutes.

To analyze our survey data, we performed Mann-Whitney tests. We compared the following audience sets: 1) AI/Tell Before vs. Non-AI/Tell Before, 2) Non-AI/Tell Before vs. Non-AI/Tell After, 3) AI/Tell Before vs. AI/Tell After, and 4) AI/Tell After vs. Non-AI/Tell After. Because we added six additional Likert scale questions to the surveys in the "Tell After" conditions, we only included the 29 baseline questions in our Mann-Whitney tests for all but the last comparative case (AI, Tell After vs. Non-AI, Tell After). In the "Tell After" condition, we included the six additional questions as dependent variables in our Mann-Whitney test.

\subsection{Procedure}
This research was approved by our institution's IRB. The procedure across all performances was the same except for the time we informed participants of technology. For the "Tell Before" performances, we informed participants before the performance started. For the "Tell After" performances, we informed them after they had taken the survey. 

Audience members were led into the studio where they sat around the stage. For five minutes, audience members of the "Tell Before" performances heard technology's role in what they would see. The performance then began and lasted for 15 minutes. Afterwards, informed consent documents for the study were distributed; interested individuals who verbally consented to the researcher could scan a QR code to the survey, which took approximately 10 minutes. After the survey, audience members of the "Tell After" performances heard how technology was used.

\section{Results}
\label{sec:results}
There was statistical significance (confidence interval = 95\%) for a selection of questions after conducting Mann-Whitney tests for the four performance sets where one independent variable was held constant; these are shown in Table \ref{table:stats} in Appendix \ref{ase:appendixresults}.

Statistically significant survey response differences between the Non-AI performances indicated that those told about technology before were watching for relationships between the dynamic visuals and the dancer compared to those told after. Comparing the AI performance responses, those told before the performance more greatly agreed with the statement that the visuals appeared random. Those told about the AI after the survey responded significantly higher on statements related to the artistic merit of the work.

\section{Discussion and Conclusion} \label{sec:discussionfuturework}
For those who experienced the Non-AI condition, the audience told before rated higher on survey questions pertaining to thinking about patterns in the visuals. This audience group also thought the visuals were more distracting than the audience told after. These results were in line with work on the effects of learned value on attentional capture \cite{anderson2011learned}. Once the audience was aware of the visuals, they became more salient to them. 

For those who experienced the AI condition, the audience told after the survey ended rated significantly higher on survey questions pertaining to the artistic merit of the visuals. The audience told before the performance rated significantly higher on the survey question: "The projected visuals appeared random." Our results suggest that viewers' awareness of how a creative work is produced influences their perception of the work's artistic value. Participants in the AI/Tell After condition heard how AI was used, but they didn't hear how AI works, which diverges from the classical emphasis in XAI on explaining how AI algorithms work. As methods in developing the best GenAI tools are becoming somewhat of a trade secret and are virtually impossible to reverse engineer the outputs of, the present study makes a case to the XAI community for a greater focus on transparency of the presence and power of GenAI in the arts and beyond. It could be valuable to design a future study integrating these two approaches towards XAI: knowledge of AI's presence and knowledge of AI's mechanics.

When comparing the results for the "Tell After" audiences, those who experienced the AI performance rated higher on survey questions related to their curiosity. Those who experienced the Non-AI performance rated higher on seeing the complementary nature of the sound. These results align with research on the behavior of GenAI and art interpretation. GenAI is known to make mistakes \cite{kim2025chatgpt, prather2024widening} and prompters lack fine-grained control in driving the output \cite{oppenlaender2024prompting, liu2022design}. Those lacking context when viewing art can perceive any product made by an artist as intentional \cite{bayles2023art}. Thus, an audience can attribute meaning to artistic products of GenAI that were unintended by the prompter. 

Participants who experienced the AI/Tell Before performance were more curious as to why certain artistic decisions were made. These participants may have been searching for artistic intention in the AI performance which could have been more apparent in the non-AI performance. We see from the higher audience ratings on the complementary nature of the sound in the non-AI performance that participants could draw relational conclusions between the elements of the non-AI performance more easily than the AI performance. This could have contributed to different judgments on the creative merits of the performances.

Finally, we utilized both GenAI (a large language model) and traditional AI (a manual machine learning tool to train and run simple neural network models) in developing our AI performances. These technologies vary greatly and may require different explainability approaches. The XAI community may benefit from isolating these AI tools to test new methodologies in XAI.

Though we took precautions in the present study, we acknowledge that there is always a possibility for bias. We did experience some technical issues while implementing the performance. In the present study we collected feedback from audience members to understand how their impressions of an artistic work changed when we (RQ1) withheld information regarding how the artwork was created, and (RQ2) used different types of technology. Using audience surveys, we found trends indicating that withholding information on how an artwork was produced can impact audience members' evaluation of the work.


\bibliographystyle{ACM-Reference-Format}
\bibliography{sample-base}


\appendix

\section{Performance Development Imagery}
\label{ase:appendixphotos}

\begin{center}
\begin{minipage}{0.5\textwidth}
  \centering
  \includegraphics[width=\linewidth]{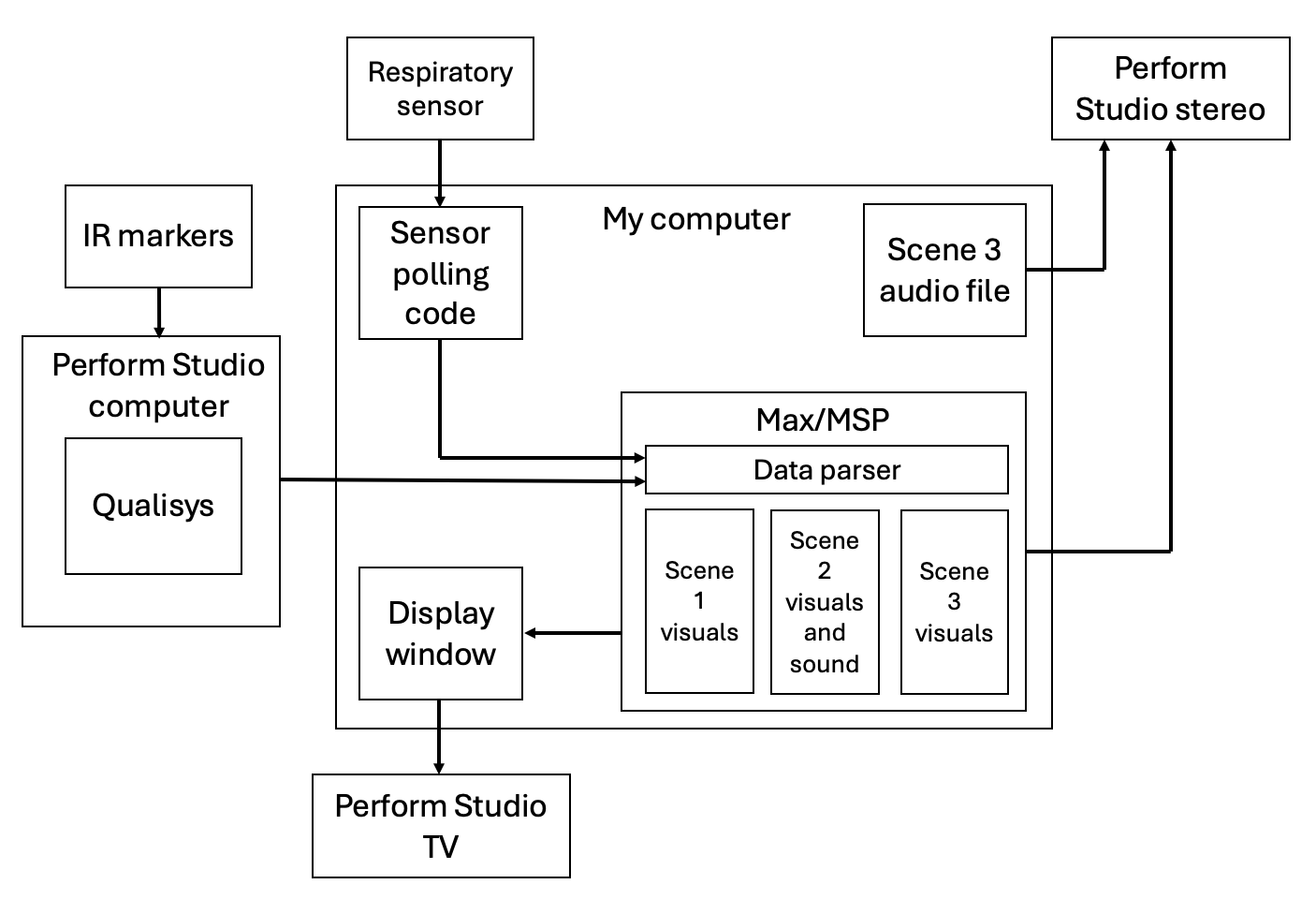}
  
  Figure 1. System architecture of the Non-AI performance version. 
  \Description{Figure 1. A diagram of all of the hardware and software components used to run the technical side of the Non-AI performance.}
  \label{fig:nonAIarch}
\end{minipage}
\end{center}

\begin{center}
\begin{minipage}{0.5\textwidth}
  \centering
  \includegraphics[width=\linewidth]{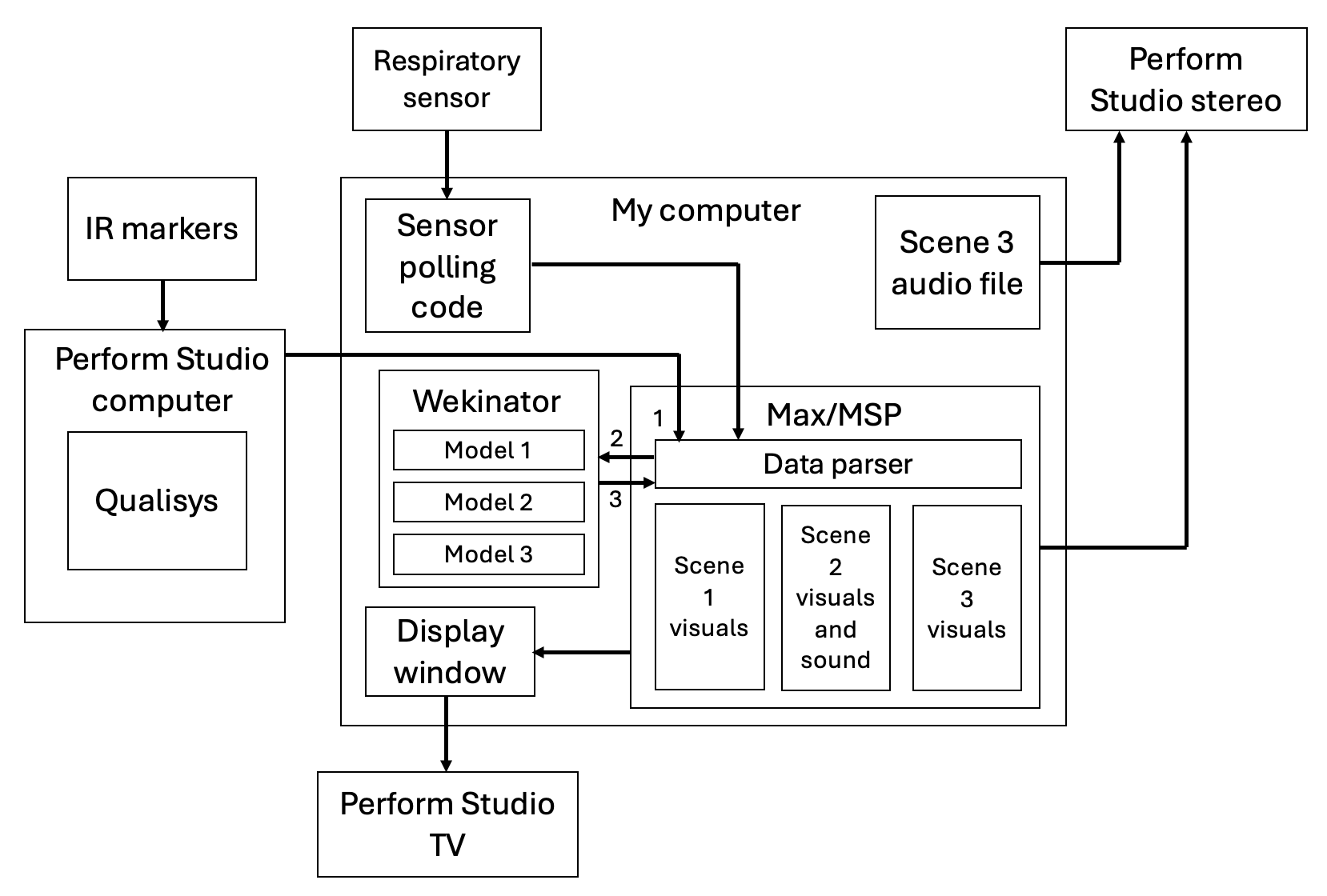}
  
  Figure 2. System architecture of the AI performance version.
  \Description{Figure 2. A diagram of all of the hardware and software components used to run the technical side of the Non-AI performance.}
  \label{fig:aiArch}
\end{minipage}
\end{center}

\begin{center}
\begin{minipage}{0.4\textwidth}
  \centering
  \includegraphics[width=.9\linewidth]{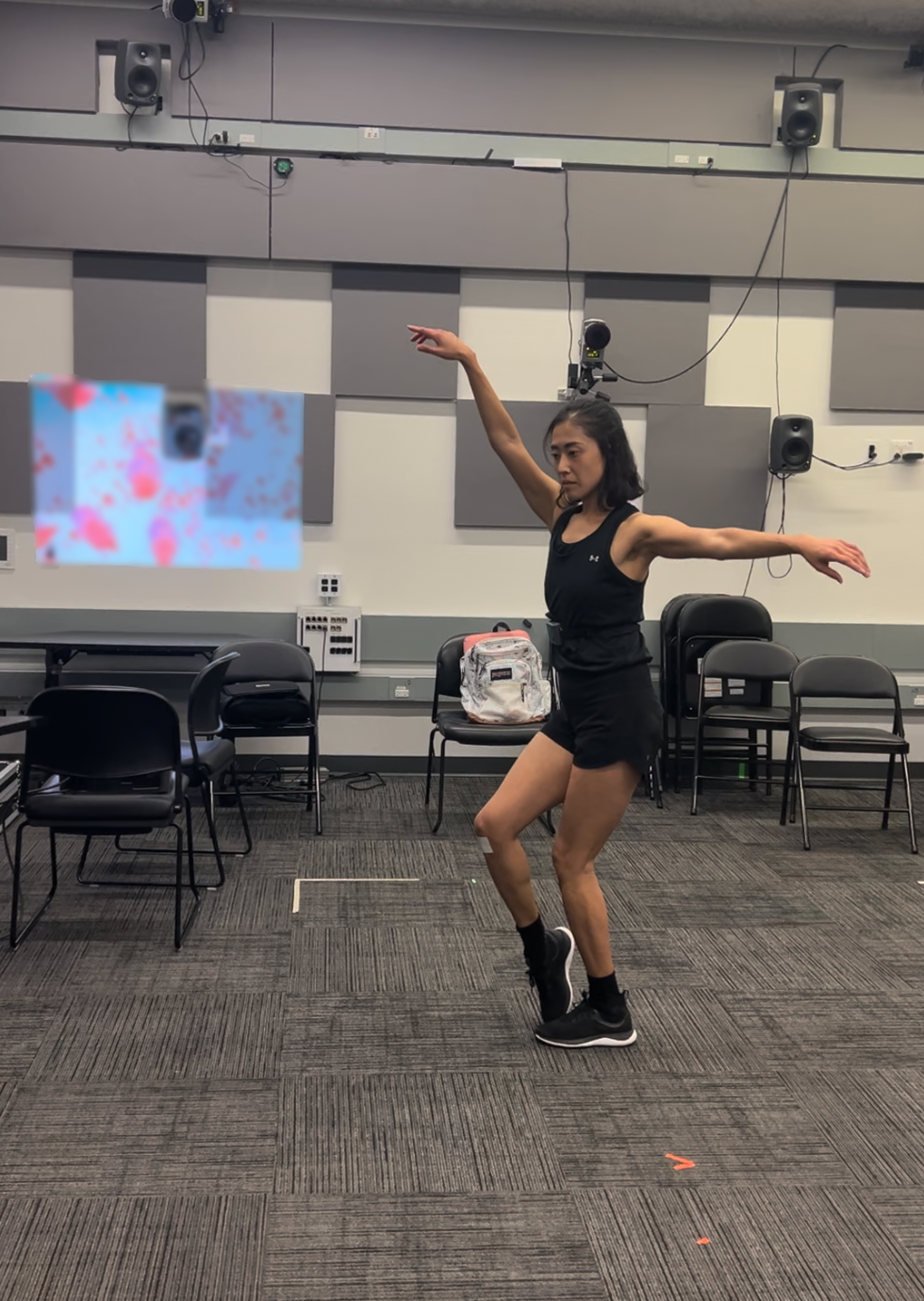}
  
  Figure 3. The dancer rehearsing during an experiment of different projection strategies.  
  \Description{Figure 3. The dancer makes an elegant pose with her arms in the air as petals are projected on the wall behind her.}
  \label{fig:rehearsal}
\end{minipage}
\end{center}

\begin{center}
\begin{minipage}{0.4\textwidth}
  \centering
  \includegraphics[width=\linewidth]{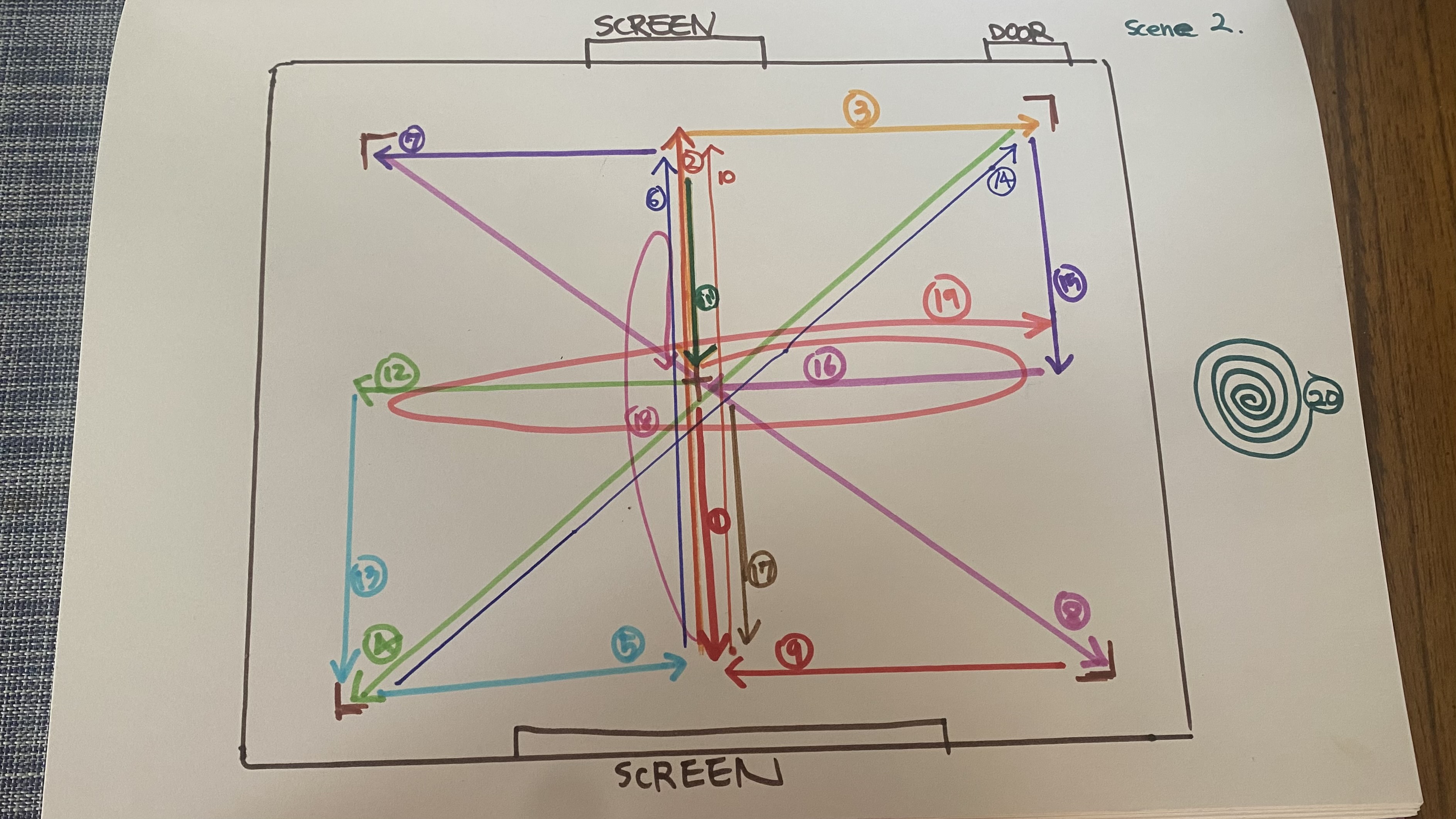}
  
  Figure 4. Outline of the choreography for Scene 2.  
  \Description{Figure 4. A sheet of paper with colored lines which are numbered, denoting where the dancer would move along the dance floor during Scene 2.}
  \label{fig:choreosketch}
\end{minipage}
\end{center}

\begin{center}
\begin{minipage}{0.35\columnwidth}
  \includegraphics[width=\linewidth]{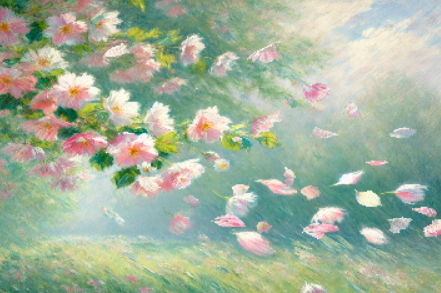} \\[0.5ex]
(a) AI Scene 1 imagery 
  \Description{Figure 5.a Light pink petals floating to the ground from a light green tree. The image looks like a painting.}
  \label{fig:fig1a} \\

  \includegraphics[width=\linewidth]{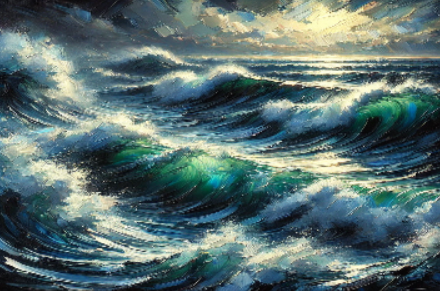} \\[0.5ex]
(b) AI Scene 2 imagery 
  \Description{Figure 5.b High waves in a stormy sea with the light of the sun shining in the distance.}
  \label{fig:fig1b} \\

  \includegraphics[width=\linewidth]{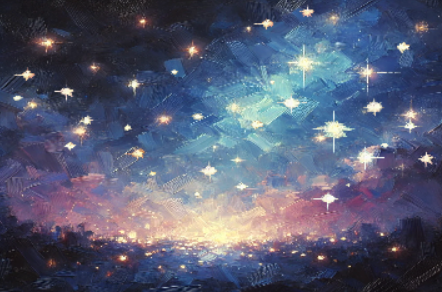}
  (c) AI Scene 3 imagery 
  \Description{Figure 5.c Stars glowing white behind a pink, yellow and blue night sky. The image looks like a painting.}
  \label{fig:fig1c} \\

  Figure 5. Imagery for each scene of the AI performance version.
\end{minipage}
\hfill
\begin{minipage}{0.6\columnwidth}
\includegraphics[width=\linewidth]{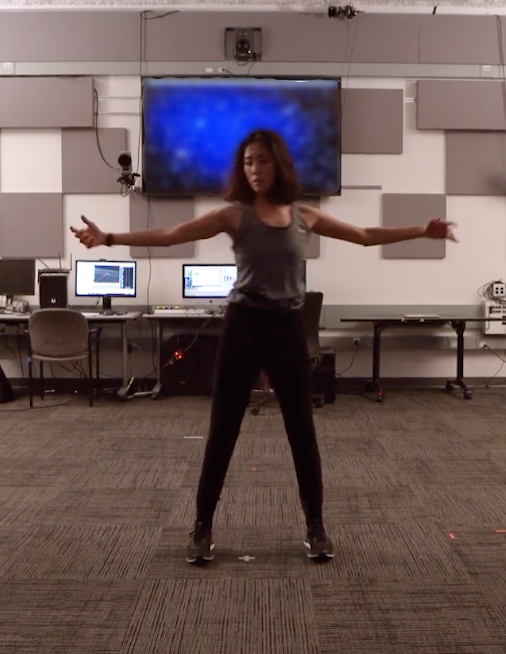}
Figure 6. The dancer during the performance.  
\Description{Figure 6. A dancer with her arms spread out in front of a TV screen displaying stars.}
\label{fig:suminsubfig}
\end{minipage}
\end{center}

\section{Research Methods}

\subsection{ChatGPT Prompts and Used Responses}\label{ase:prompts}
\textit{Prompt for Scene 1}:

"If I had to make petals of a painting move according to the respiratory rate of a dancer how should I map this respiratory rate to the petals?"

\textit{Used Response for Scene 1}:

"...the petals could gently rotate based on changes in the breathing depth and they could also rise and fall based on the breathing cycles."

\textit{Prompts for Scene 2}:
\begin{itemize}
    \item "If I had to make waves move according to a continuous stream of three probabilities representing how close a dancer's movements are to one of three movements, how should I map the three continuous probabilities to the wave movements?"
    \item "If I had to make melodious sound according to a continuous stream of three probabilities representing how close a dancer's movements are to one of three movements, how should I map the three continuous probabilities to sound? Please give me details on pitch and tempo."
    \item "What pentatonic, minor, and chromatic scales should I use?"
    \item "What octaves should I use for the C major pentatonic scale, the A minor scale, and the chromatic scale from D to A if I will be using them together? What instrument should I use to complement graceful dancing?"
\end{itemize}
\textit{Used Response for Scene 2}:

"...assign each probability a scale or a set of pitches... Movement A can map to a pentatonic scale.. Movement B can use a minor scale, and Movement C can use a chromatic scale." ChatGPT further suggested we use a C major pentatonic scale, an A minor scale, and a chromatic scale from D to A for each movement respectively.

\textit{Prompt for Scene 3}:

"If I had to make stars in a night sky change according to a continuous stream of data representing a dancer's respiratory rate how should I map this respiratory rate data stream to the stars?" 

\textit{Used Responses for Scene 3}:

\begin{itemize}
    \item "...assign one probability [of a classified dance move currently happening] to control the height or amplitude of the waves... another probability could influence the wave's direction or angle."
    \item "...you could also experiment with the mappings to see what creates the most engaging visual experience." We took this liberty by mapping the third dance move's resemblance probability to the horizontal theta cosine offset of the wave. 
    \item "...make the stars move more rapidly across the sky, suggesting heightened energy for a higher respiration rate. Their movement could be subtle or more dynamic, depending on how you want to emphasize the breathing. For a slower respiratory rate, the stars can move more slowly or remain still for a tranquil effect."
\end{itemize}

\subsection{Audience Survey}\label{ase:audiencesurvey}
All survey questions were asked of all audience groups unless otherwise noted. All questions were answered on a Likert scale from (1 - "Not at all" to 5 - "Absolutely") unless otherwise noted.

\textbf{Performance Related Questions - Visuals}
\textit{Throughout the performance, we projected various visuals as the dancer performed. Please answer the following questions based on the performance you just watched.}

\begin{itemize}
    \item I paid attention to the visuals. 
    \item The projected visuals were pleasing. 
    \item The projected visuals complemented the performance. 
    \item The projected visuals enhanced the performance. 
    \item The projected visuals made sense in the context of the whole performance.
    \item The projected visuals demonstrated artistic merit. 
    \item The projected visuals were creative. 
    \item The projected visuals seemed meaningful.
    \item The projected visuals were distracting.
    \item The projected visuals appeared random.
    \item I think I would feel the same way about the performance if it didn't have visuals.
    \item I tried to see if there was a pattern in the visuals.
    \item I wondered what caused the visuals to change. AFTER
    \item Which of the following do you feel to be the most accurate representation of the performance? (Question asked only to those in the "Tell After" condition) (Likert scale ranged from: 1 - "The visuals mostly seemed to guide the dancer." to 5 - "The dancer mostly seemed to guide the visuals.")
    \item If I knew exactly what the visuals represented, I think I would enjoy the performance more. (Question asked only to those in the "Tell After" condition)
    \item I thought about how the visuals were created. 
    \item I wondered why the visuals were chosen.
    \item I tried to make a connection between what the dancer was doing and the visuals.
\end{itemize}

\textbf{Performance Related Questions - Sound}
\textit{In part 2, when we showed the wave visuals, there were some sounds that you heard during the performance. Please answer the following questions about this audio you heard.}

\begin{itemize}
    \item I didn't pay attention to the sound in part 2.
    \item The sound in part 2 was pleasant. 
    \item The sound in part 2 complemented the performance.
    \item The sound in part 2 enhanced the performance.
    \item The sound in part 2 fit with the other sounds in the performance. 
    \item The sound in part 2 demonstrated artistic merit.
    \item The sound in part 2 was creative. 
    \item The sound in part 2 was meaningful. 
    \item The sound in part 2 was distracting. 
    \item The sound in part 2 seemed random. 
    \item I think I would feel the same way about the performance if it didn't have sound in part 2.
    \item I tried to see if there was a pattern in sound in part 2.
    \item I was curious about why the sound in part 2 changed. (Question asked only to those in the "Tell After" condition)
    \item Which of the following do you believe to be the most accurate representation of part 2 of the performance specifically? (Question asked only to those in the "Tell After" condition) (Likert scale ranged from: 1 - "The sound mostly seemed to guide the dancer." to 5 - "The dancer mostly seemed to guide the sound.")
    \item If I knew exactly why the sounds were chosen, I think I would enjoy the performance more. (Question asked only to those in the "Tell After" condition)
    \item I thought about how the sound in part 2 was created. 
    \item I wondered why the sound in part 2 was chosen. 
    \item I tried to make a connection between what the dancer was doing and the sound in part 2.
    \item I noticed a mapping between the dancer's movements and the sound in part 2. (Question asked only to those in the "Tell After" condition) (Response choices were "Yes," "No," "I don't recall.")
    \item Can you briefly describe what you think the mapping between the dancer's movements and the sound in part 2 might be? (Question asked only to those in the "Tell After" condition) (Answers were written responses)
\end{itemize}

\section{Quantitative Results}
\label{ase:appendixresults}
\clearpage

\begin{table*}[h]
\centering
\begin{tabular}{|p{2.3cm}|p{14.7cm}|}
    \hline
    Performance Comparison & Statistically Significant Survey Response Differences \\
    \hline

    Both told before, comparing technology type & 
    $\bullet$ The projected visuals enhanced the performance (\textit{Z} = -2.667, \textit{p} = .008, the Non-AI performance rated higher, [AI version: \textit{M} = 2.00, \textit{SD} = .739; Non-AI version: \textit{M} = 3.25, \textit{SD} = .886]). \newline
    $\bullet$ The projected visuals were distracting (\textit{Z} = -2.025, \textit{p} = .043, the Non-AI performance rated higher, [AI version: \textit{M} = 2.00, \textit{SD} = 1.348; Non-AI version: \textit{M} = 3.13, \textit{SD} = .991]). \newline
    $\bullet$ The projected visuals appeared random (\textit{Z} = -2.130, \textit{p} = .033, the AI performance rated higher, [AI version: \textit{M} = 3.67, \textit{SD} = 1.073; Non-AI version: \textit{M} = 2.62, \textit{SD} = .916]). 
    \\
    \hline

    Both Non-AI, comparing time told & 
    $\bullet$ The projected visuals were distracting (\textit{Z} = -3.333, \textit{p} = <.001, the "Tell Before" performance rated higher, [Tell before: \textit{M} = 3.13, \textit{SD} = .991; Tell after: \textit{M} = 1.30, \textit{SD} = .483]). \newline
    $\bullet$ I tried to see if there was a pattern in the visuals (\textit{Z} = -2.789, \textit{p} = .005, the "Tell Before" performance rated higher, [Tell before: \textit{M} = 4.88, \textit{SD} = .354; Tell after: \textit{M} = 3.60, \textit{SD} = 1.075]). \newline
    $\bullet$ I thought about how the visuals were created. (\textit{Z} = -3.180, \textit{p} = .001, the "Tell Before" performance rated higher, [Tell before: \textit{M} = 4.50, \textit{SD} = .756; Tell after: \textit{M} = 2.30, \textit{SD} = 1.160]). \newline
    $\bullet$ ]). I tried to make a connection between what the dancer was doing and the visuals. (\textit{Z} = -2.049, \textit{p} = .040, the "Tell Before" performance rated higher, [Tell before: \textit{M} = 4.88, \textit{SD} = .354; Tell after: \textit{M} = 3.90, \textit{SD} = 1.370]). 
    \\
    \hline

    Both AI, comparing time told & 
    $\bullet$ The projected visuals complemented the performance (\textit{Z} = -2.411, \textit{p} = .016, the "Tell After" performance rated higher, [Tell before: \textit{M} = 2.33, \textit{SD} = .778; Tell after: \textit{M} = 3.33, \textit{SD} = 1.00]). \newline
    $\bullet$ The projected visuals enhanced the performance (\textit{Z} = -2.007, \textit{p} = .045, the "Tell After" performance rated higher, [Tell before: \textit{M} = 2.00, \textit{SD} = .739; Tell after: \textit{M} = 2.89, \textit{SD} = 1.054]).  \newline
    $\bullet$ The projected visuals demonstrated artistic merit (\textit{Z} = -2.501, \textit{p} = .012, the "Tell After" performance rated higher, [Tell before: \textit{M} = 2.83, \textit{SD} = 1.030; Tell after: \textit{M} = 4.11, \textit{SD} = 1.054]). \newline
    $\bullet$ Being informed about the production of a piece of art would impact its monetary valueThe projected visuals appeared random (\textit{Z} = -2.698, \textit{p} = .007, the "Tell Before" performance rated higher, [Tell before: \textit{M} = 3.67, \textit{SD} = 1.073; Tell after: \textit{M} = 2.22, \textit{SD} = .972]).  
    \\
    \hline

    Both told after, comparing technology type & 
    $\bullet$ The projected visuals were distracting (\textit{Z} = -2.008, \textit{p} = .045, the AI performance rated higher, [AI version: \textit{M} = 2.56, \textit{SD} = 1.509; Non-AI version: \textit{M} = 1.30, \textit{SD} = .483]). \newline
    $\bullet$ I thought about how the visuals were created (\textit{Z} = -2.795, \textit{p} = .005, the AI performance rated higher, [AI version: \textit{M} = 4.11, \textit{SD} = 1.054; Non-AI version: \textit{M} = 2.30, \textit{SD} = 1.160]). \newline
    $\bullet$ I wondered why the visuals were chosen (\textit{Z} = -2.162, \textit{p} = .031, the AI performance rated higher, [AI version: \textit{M} = 4.44, \textit{SD} = .726; Non-AI version: \textit{M} = 3.50, \textit{SD} = .972]). \newline
    $\bullet$ The sound in part 2 complemented the performance (\textit{Z} = -2.061, \textit{p} = .039, the Non-AI performance rated higher, [AI version: \textit{M} = 3.56, \textit{SD} = .882; Non-AI version: \textit{M} = 4.30, \textit{SD} = .483]). \newline
    $\bullet$ The sound in part 2 enhanced the performance (\textit{Z} = -2.253, \textit{p} = .024, the Non-AI performance rated higher, [AI version: \textit{M} = 3.78, \textit{SD} = .667; Non-AI version: \textit{M} = 4.50, \textit{SD} = .527]). \newline
    $\bullet$ The sound in part 2 fit with the other sounds in the performance (\textit{Z} = -1.973, \textit{p} = .048, the Non-AI performance rated higher, [AI version: \textit{M} = 2.44, \textit{SD} = .882; Non-AI version: \textit{M} = 3.50, \textit{SD} = 1.269]).  \newline
    $\bullet$ I wondered what caused the visuals to change (\textit{Z} = -2.201, \textit{p} = .028, the AI performance rated higher, [AI version: \textit{M} = 4.33, \textit{SD} = .707; Non-AI version: \textit{M} = 3.20, \textit{SD} = 1.135]).  
    \\
    \hline

\end{tabular}
\caption{\normalfont Statistically significant survey responses rendered from Mann-Whitney tests}
\Description{Table 1. No alternative text is needed in this case because the caption and table contents include all the necessary information.}
\label{table:stats}
\end{table*}

\clearpage

\end{document}